\documentclass[twocolumn]{article}
\usepackage[margin=0.6in,columnsep=0.15in]{geometry}
\usepackage[utf8]{inputenc}

\usepackage[usenames,dvipsnames]{xcolor}
\usepackage[colorlinks,citecolor=Gray,urlcolor=Gray,linkcolor=Blue]{hyperref}

\usepackage{graphicx}
\usepackage{mathtools}
\usepackage{booktabs}
\usepackage{tabularx}
\usepackage{tikz}
\usepackage{siunitx}
\usepackage{dcolumn}

\usepackage[lcgreekalpha,upint]{stix}

\usepackage[tiny]{titlesec}
\usepackage[inline]{enumitem}

\usepackage[normal]{caption}
\DeclareCaptionLabelSeparator{pipe}{ $|$ }
\usepackage{authblk}

\usepackage[colon]{natbib}
\usepackage{doi}

\title{\Large Deep neural network solution of the electronic Schrödinger
equation}

\author[1,2,*]{Jan Hermann}
\author[1]{Zeno Schätzle}
\author[1,3,4,*]{Frank Noé}
\affil[1]{FU Berlin, Department of Mathematics and Computer Science,
Arnimallee 6, 14195 Berlin, Germany}
\affil[2]{TU Berlin, Machine Learning Group, Marchstr. 23, 10587 Berlin,
Germany}
\affil[3]{FU Berlin, Department of Physics, Arnimallee 14, 14195 Berlin,
Germany}
\affil[4]{Rice University, Department of Chemistry, Houston, TX 77005,
USA}

\date{}

\makeatletter
\def\blfootnote{\xdef\@thefnmark{}\@footnotetext}
\makeatother

\begin{document}

\twocolumn[{%
  \maketitle
  \vspace{-3em}
  \begin{center}
  \begin{minipage}{0.85\linewidth}
    \small
    \paragraph{Abstract}
    \textcolor{RubineRed}{[New and updated results were published in Nature Chemistry, doi:\href{http://doi.org/10.1038/s41557-020-0544-y}{\textcolor{RubineRed}{10.1038/s41557-020-0544-y}}.]}
    The electronic Schrödinger equation describes fundamental properties
    of molecules and materials, but can only be solved analytically for
    the hydrogen atom. The numerically exact full
    configuration-interaction method is exponentially expensive in the
    number of electrons. Quantum Monte Carlo is a possible way out: it
    scales well to large molecules, can be parallelized, and its
    accuracy has, as yet, only been limited by the flexibility of the
    used wave function ansatz. Here we propose PauliNet, a deep-learning
    wave function ansatz that achieves nearly exact solutions of the
    electronic Schrödinger equation. PauliNet has a multireference
    Hartree--Fock solution built in as a baseline, incorporates the
    physics of valid wave functions, and is trained using variational
    quantum Monte Carlo (VMC). PauliNet outperforms comparable
    state-of-the-art VMC ansatzes for atoms, diatomic molecules and a
    strongly-correlated hydrogen chain by a margin and is yet
    computationally efficient. We anticipate that thanks to the
    favourable scaling with system size, this method may become a new
    leading method for highly accurate electronic-structure calculations
    on medium-sized molecular systems.
  \end{minipage}
  \end{center}
  \vspace{1em}
}]

\blfootnote{$^*$Emails: jan.hermann@fu-berlin.de, frank.noe@fu-berlin.de}%



\hyphenation{Schrö-din-ger}

\section{Introduction}  


A solution of the time-independent electronic Schrödinger equation of a given atomic system provides, in principle, full access to its chemical properties.
This equation can be solved analytically only for an isolated hydrogen atom, but solid-state physics and quantum chemistry have been remarkably successful in developing numerical approximation methods \citep{Piela14}.
For small molecules containing up to a few tens of electrons, methods based on the configuration interaction and the closely related coupled cluster approaches or the multideterminant quantum Monte Carlo (QMC) can reach impressive accuracy of up to six significant digits in the total electronic energy \citep{MoralesJCTC12}.

Unfortunately, the computational cost of such high-accuracy methods increases as $N^7$ or worse with the number of electrons, $N$, making them unworkable for most molecules or materials of practical interest.
Computationally less demanding methods, such as the density functional theory, can scale to larger molecules, but at the price of limited accuracy.
Fundamentally, high-accuracy methods scale unfavorably with $N$ because the many-body electronic problem is \emph{non-polynomial hard} due to the Pauli exclusion principle \citep{TroyerPRL05}.
The tradeoff between accuracy and computational cost is apparent, when considering that most quantum chemistry methods represent electronic wave functions by linear combinations of Slater matrix determinants.
A Slater matrix is constructed by selecting $N$ out of $M>N$ molecular orbitals, and assigning $N$ electrons to them, resulting in a combinatorial growth of all possible matrices with system size.


\begin{figure}[t]
\centering
\includegraphics{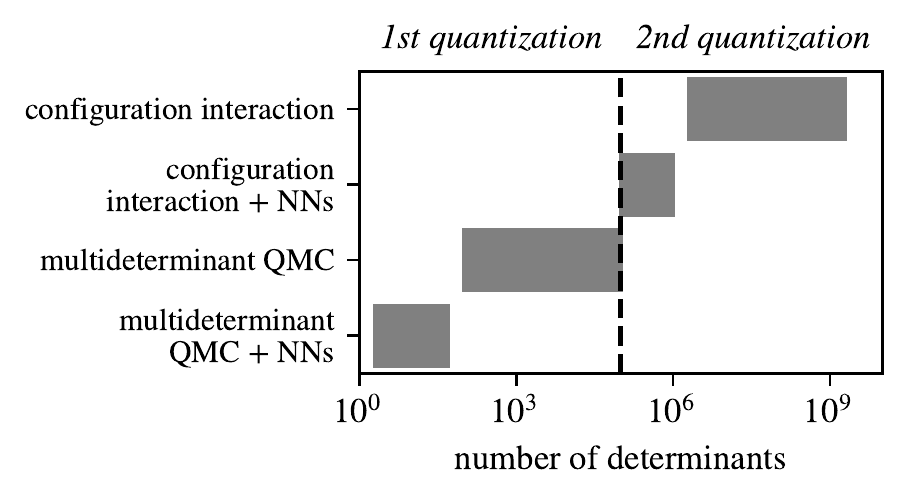}
\caption{\textbf{Typical numbers of Slater determinants required for high-accuracy quantum chemistry methods}.
The multi-determinant quantum Monte Carlo (QMC) in combination with neural networks (NNs) is the approach developed in this work.
The numbers correspond to highly accurate calculations on atomic systems with at most a few tens of electrons.
}\label{fig:ndets}
\end{figure}


The Slater determinants have different roles in different quantum chemistry methods (Figure~\ref{fig:ndets}).
In the configuration interaction and coupled cluster approaches, the electronic problem is solved entirely in the basis of the determinants (second quantization), and as such their number in typical applications is the largest, as the determinant expansion must recover all many-body interactions that are missing in individual determinants \citep{Shavitt09}.
Stochastic methods that sample over vast determinant spaces have been developed  \citep{BoothJCP09,ThomPRL10}, but the underlying scaling trap persists nevertheless.
Recent work by \citet{ChooNC20} suggests that the number of required determinants can be reduced with the use of neural networks, but whether this would also reduce the scaling issue has yet to be demonstrated.

Conventional QMC methods \citep{FoulkesRMP01,NeedsJPCM10,AustinCR12} solve the electronic problem in real space (first quantization), and treat a large part of the correlation in the electronic motion explicitly, which greatly reduces the number of required determinants.
Standard QMC variants are still practical for system with hundreds of electrons, such as supramolecular complexes \citep{AmbrosettiJPCL14} and molecular crystals \citep{ZenPNAS18}.
Even though only a single determinant can be typically afforded for these larger systems, QMC typically outperforms other electronic-structure methods.
However, for small systems, where second-quantized approaches are applicable, standard QMC methods need to use at least hundreds of determinants to be competitive, and this number increases rapidly with $N$.
The large number of determinants is necessary to accurately represent the \emph{nodal surface} of the wave function, which the real-space QMC techniques are unable to improve.
A key development facilitating the progress in this work is the real-space \emph{backflow} technique \citep{LopezRiosPRE06}.
The idea of backflow is to transform the electron positions into pseudoparticles, each of which depends on all electron positions, and then inserting these pseudoparticles into the Slater determinants, leading to an improved nodal surface.
While the traditional backflow does not reach the accuracy of the large determinant expansions and does not generalize well to larger systems, \citet{LuoPRL19} recently showed that representing the backflow with a neural network is a powerful generalization.


Machine learning has had significant impact on quantum chemistry, especially in the case of supervised learning and prediction of electronic energies \citep{BehlerPRL07,RuppPRL12,ChmielaSA17,BartokSA17,SmithCS17,SchuttJCP18,FaberJCP18,WelbornJCTC18}, electron densities \citep{GrisafiACS19}, and molecular orbitals \citep{SchuttNC19}.
This approach entirely avoids the solution of the Schrödinger equation, at the price of requiring datasets of preexisting solutions, obtained for instance by density functional theory or the coupled cluster method.

In contrast, the direct representation of highly correlated wave functions with neural networks and their unsupervised training via the variational principle, first proposed by \citet{CarleoS17} for discrete spin lattice systems, is an ab-initio approach that requires no preexisting data and has no fundamental limits to its accuracy.
It is motivated by the fact that neural networks are universal function approximators, and could therefore provide more efficient means for approximating the exponentially scaling wave function complexity of many-body systems.
The initial attempts on lattice systems were later generalized to bosons in real space \citep{SaitoJPSJ18}, and even electrons in real space \citep{HanJCP19}, but the latter approach does not use a wave function ansatz in the form of a Slater determinant, and perhaps for that reason does not reach the accuracy of even the baseline Hartree--Fock method for some systems.

In this work, we develop PauliNet, a deep learning QMC approach that replaces existing ad-hoc functional forms used in the standard Jastrow factor and backflow transformation with more powerful deep neural network representations.
Besides the sheer gain in expressive power, our neural network architecture is specifically designed to encode the physics of valid wave functions and incorporates a multireference Hartree-Fock method as a baseline.
These physically motivated choices are essential in order to obtain a method that is not only highly accurate, but also converges robustly, and still maintains relatively high computational efficiency.
On a small set of test systems, we show that our neural network ansatz requires only a few determinants to surpass the accuracy of state-of-the-art single-determinant wave function ansatzes by a margin, and approaches the accuracy of multi-determinant QMC methods that use orders of magnitude more determinants.
As a result, our method has the asymptotic scaling of $N^4$, and we expect that it will be feasible to apply it to much larger systems than is currently possible with existing highly accurate methods.

The parallel work of \citet{Pfau19} follows the same basic idea as ours, but differs in several significant aspects.
Their architecture does not encode any physical knowledge about wave functions besides the essential antisymmetry, which is compensated by a much larger number of optimized parameters.
Perhaps as a result of that, their ansatz achieves somewhat higher accuracy at the cost of being about two orders of magnitude more computationally expensive than ours.

\section{Results}  

\subsection{Deep neural network electronic wave function ansatz}  

\begin{figure}[t]
\centering
\includegraphics[width=\columnwidth]{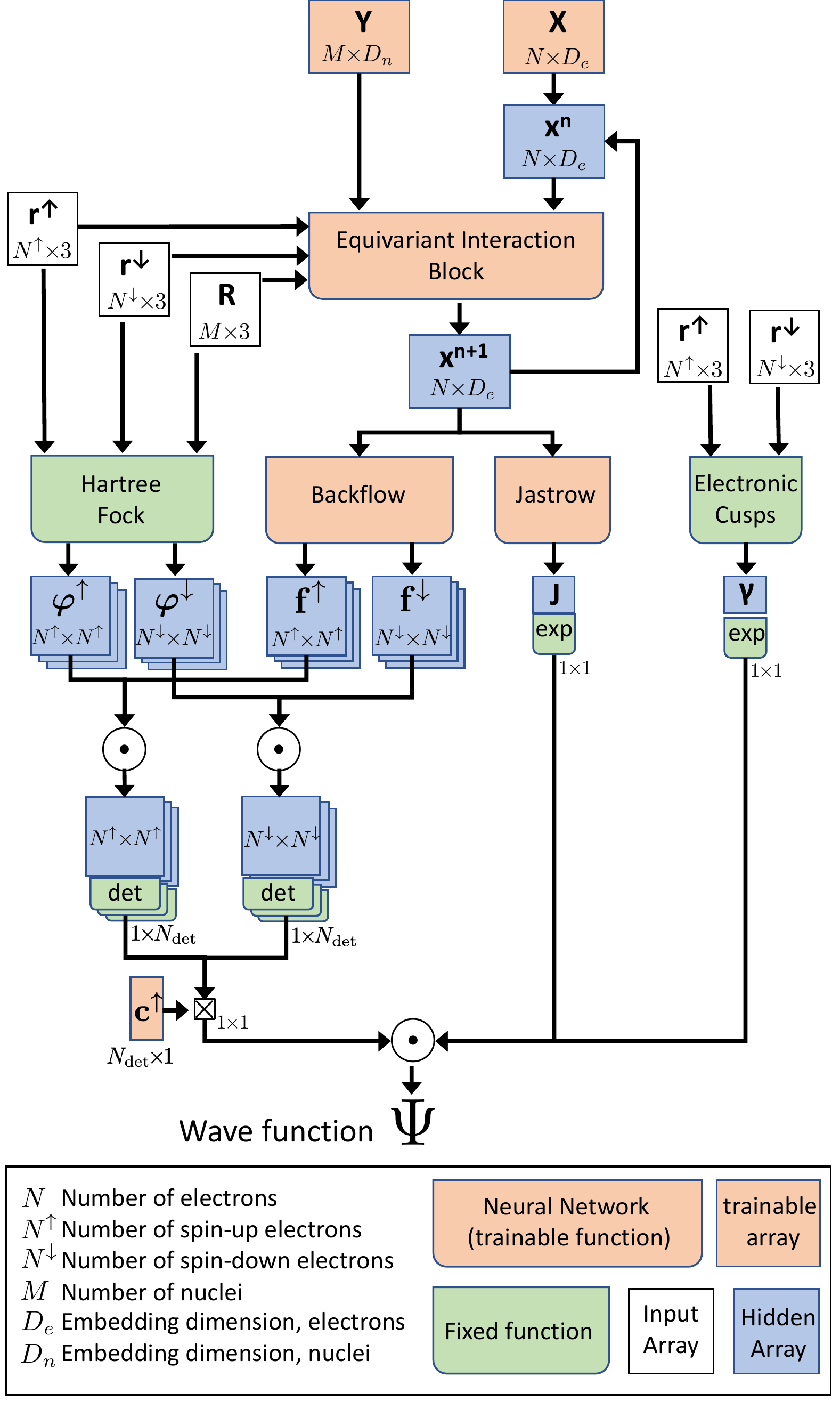}
\caption{\textbf{Information flow in the PauliNet ansatz architecture.}
}\label{fig:scheme}
\end{figure}


At the core of our deep learning approach to the electronic Schrödinger equation is a wave function ansatz, dubbed PauliNet, which incorporates both the well-established essential physics of electronic wave functions---Slater determinants, multi-determinant expansion, Jastrow factor, backflow transformation, cusp conditions---as well as deep neural networks capable of encoding the complex features of the electronic motion in heterogeneous molecular systems.
Our proposed trial wave function is of the multi-determinant Slater--Jastrow--backflow type \citep{BrownJCP07}, where both the Jastrow factor, $J$, and the backflow, $\mathbf f$, are represented by deep neural networks (DNNs) with trainable parameters $\boldsymbol\theta$ (Figure~\ref{fig:scheme}),
\begin{equation}
\begin{gathered}
\psi_{\boldsymbol\theta}(\mathbf r)
  =\mathrm e^{\gamma(\mathbf r)+J_{\boldsymbol\theta}(\mathbf r)}
  \sum_p c_p
  \det[\tilde\varphi_{\boldsymbol\theta,{\mu_p}i}^\uparrow(\mathbf r)]
  \det[\tilde\varphi_{\boldsymbol\theta,{\mu_p}i}^\downarrow(\mathbf r)]
  \\
\tilde\varphi_{\mu i}(\mathbf r)
  =\varphi_\mu(\mathbf r_i)
  f_{\boldsymbol\theta,\mu i}(\mathbf r)
  \end{gathered}
  \label{eq:ansatz}
\end{equation}
While the expressiveness of PauliNet is contained in the Jastrow factor and backflow DNNs, the physics is encoded by the determinant form, the one-electron molecular orbitals, $\varphi_\mu$, and the electronic cusps, $\gamma$, in the following way:

\paragraph{Antisymmetry} Every valid electronic wave function must be antisymmetric with respect to the exchange of same-spin electrons, 
\begin{equation}
  \psi(\ldots,\mathbf r_i,\ldots,\mathbf r_j,\ldots)=-\psi(\ldots,\mathbf r_j,\ldots,\mathbf r_i,\ldots)
  \label{eq:antisymmetry}
\end{equation}
As is common in quantum chemistry, we enforce antisymmetry via matrix determinants, as determinants change sign upon exchanging any two rows or columns.

\paragraph{Hartree--Fock baseline} To ensure a good starting point for the variational optimization problem, we exploit the approximate Hartree--Fock (HF) quantum chemistry method.
Specifically, we use a multireference HF calculation with a small complete active space, and select the most dominant determinants and their orbitals based on the magnitude of the linear coefficient.
The HF-optimized one-electron molecular orbitals, $\varphi_\mu(\mathbf r)$, are then used as an input to PauliNet, and are modified during training only by the backflow transformation.
    
\paragraph{Asymptotic behavior}
Any ground-state electronic wave function obeys exact asymptotic behavior defined by the \emph{cusp conditions} as electrons approach each other and the nuclei \citep{KatoCPAM57}.
We chose to build the cusp conditions directly into the PauliNet functional form as this makes the training more efficient and stable by removing divergences from the local electronic energy.
We incorporate the nuclear cusps by modifying the molecular orbitals using the technique from \citet{MaJCP05}, and the electronic cusps by the fixed cusp function, $\gamma(\mathbf r)$.
We ensure that the trainable Jastrow factor and backflow DNNs are cuspless, so as to maintain the enforced cusp behavior (see Methods for details).


\subsection{Robust deep Jastrow factor and backflow}  %

PauliNet differs from conventional QMC ansatzes by representing the Jastrow factor and backflow functions with specialized DNNs.
To retain the antisymmetry of the wave function, as enforced by the Slater determinants, the Jastrow factor and backflow DNNs are constructed to be invariant and equivariant, respectively, with respect to the exchange of same-spin electrons, $\mathcal P_{ij}$,
\begin{equation}
J(\mathcal P_{ij}\mathbf r)=J(\mathbf r),\qquad
  \mathcal P_{ij}f_{\mu i}(\mathbf r)=f_{\mu j}(\mathcal P_{ij}\mathbf r)
\end{equation}
The Jastrow factor is a nonnegative totally symmetric function with the commonly used form, $\mathrm e^{J(\mathbf r)}$, which can encode complex electron correlations into the wave function, but cannot modify the nodal surface inherited from the determinant expansion.

We found that attempting to express the standard backflow form of nonlocal coupled coordinates of pseudoelectrons that enter the one-electron molecular orbitals with DNNs leads to a difficult optimization problem.
Instead, the PauliNet backflow has the form of multiplying the bare one-electron molecular orbitals with many-electron equivariant functions, $\mathbf f$ (Eq.~\ref{eq:ansatz}).
In combination with just a few determinants, this presents a powerful representation of the electronic nodal surface.

The requirements of invariance and equivariance with respect to permutation of particles, and the fact that particle interactions are a function of their distances, are closely related to the aim of constructing DNNs that learn potential energy functions.
PauliNet uses an adapted form of one such DNN architecture, called SchNet \citep{SchuttJCP18}.
SchNet is a graph convolution DNN that represents each particle with a vector in a high-dimensional abstract feature space, $\mathbf x_i$, which is first initialized by a trainable vector shared by all particles of the same type, and then iteratively refined by interactions with other particles through real-space trainable convolutions, $\boldsymbol\chi_{\boldsymbol\theta}$, which encode the inter-particle distances and are equivariant with respect to particle exchange,
\begin{equation}
    \mathbf x_i^{(n+1)}:=\mathbf x_i^{(n)} + \boldsymbol\chi^{(n)}_{\boldsymbol\theta}\big(\big\{\mathbf x_j^{(n)},\{\lvert\mathbf r_j-\mathbf r_k\rvert\}\big\}\big)
    \label{eq:schnet-iter}
\end{equation}
The SchNet architecture and its modifications for PauliNet are described in detail in Methods.
After a fixed number of iterations of SchNet, the final electron representations, $\mathbf x_i^{(L)}$, which now encode complex many-body electron correlations, are used as an input to two trainable functions, $\eta_{\boldsymbol\theta}$, $\boldsymbol\kappa_{\boldsymbol\theta}$, which return the Jastrow factor and the backflow, respectively.
\begin{equation}
  J:=\eta_{\boldsymbol\theta} \big(\textstyle\sum_i\mathbf x_i^{(L)}\big),\qquad
  \mathbf f_i:=\boldsymbol\kappa_{\boldsymbol\theta}\big(\mathbf x_i^{(L)}\big),
\end{equation}
%
Since the feature vectors, $\mathbf x_i^{(n)}$, are equivariant with respect to electron exchange at each iteration, so are the backflow vectors, $\mathbf f_i$.
As a result, the Slater determinants in PauliNet produce an antisymmetric wave function.
Furthermore, $J$ is by construction invariant with respect to exchanges of electrons and therefore a symmetric function that maintains this antisymmetry.

\subsection{Approaching exact solution with few determinants}  

\begin{table*}[t]
\centering
\caption{\textbf{Ground-state energies of five test systems obtained by four different methods.}
}\label{tab:energies}
\begin{tabular}{l*{4}{D{.}{.}{2.3}}*{2}{D{.}{.}{3.7}}}
\toprule
& \multicolumn4c{correlation energy} & \multicolumn2c{$E/\si{\hartree}$} \\
system & \multicolumn1c{SD-VMC} & \multicolumn1c{SD-DMC} & \multicolumn1c{DeepWF} & \multicolumn2c{PauliNet$^g$} & \multicolumn1c{exact} \\
\midrule
H$_2$    &          &          & 98.4\%               & 99.99\% & -\mathbf{1}.\mathbf{17447}(2)  & -1.17447^a  \\
LiH    & 91.5\%^b & 99.7\%^b & \multicolumn1c{$^c$} & 99.3\%  & -\mathbf{8}.\mathbf{070}0(2)   & -8.070548^d \\
Be     & 61.6\%^e & 89.2\%^e & 43.6\%               & 99.94\% & -\mathbf{14}.\mathbf{667}4(3)  & -14.66736^e \\
B      & 60.0\%^e & 88.3\%^e & \multicolumn1c{$^c$} & 97.3\%  & -\mathbf{24}.\mathbf{65}06(11) & -24.65391^e \\
H$_{10}$ &          &          & 63.8\%               & 98.0\%  & -\mathbf{5}.\mathbf{66}02(7)   & -5.6655^f   \\
\bottomrule
\end{tabular}

\footnotesize
\begin{minipage}{0.65\linewidth}
$^a$\citet{KolosJCP65}.
$^b$\citet{CasalegnoJCP03}.
$^c$For LiH and B, DeepWF does not reach the accuracy of the HF method.
$^d$\citet{CencekCPL00}.
$^e$\citet{BrownJCP07}.
$^f$\citet{MottaPRX17}.
$^g$ Bold digits match the exact energy.
\end{minipage}
\end{table*}

\begin{figure}[t!]
\centering
\includegraphics{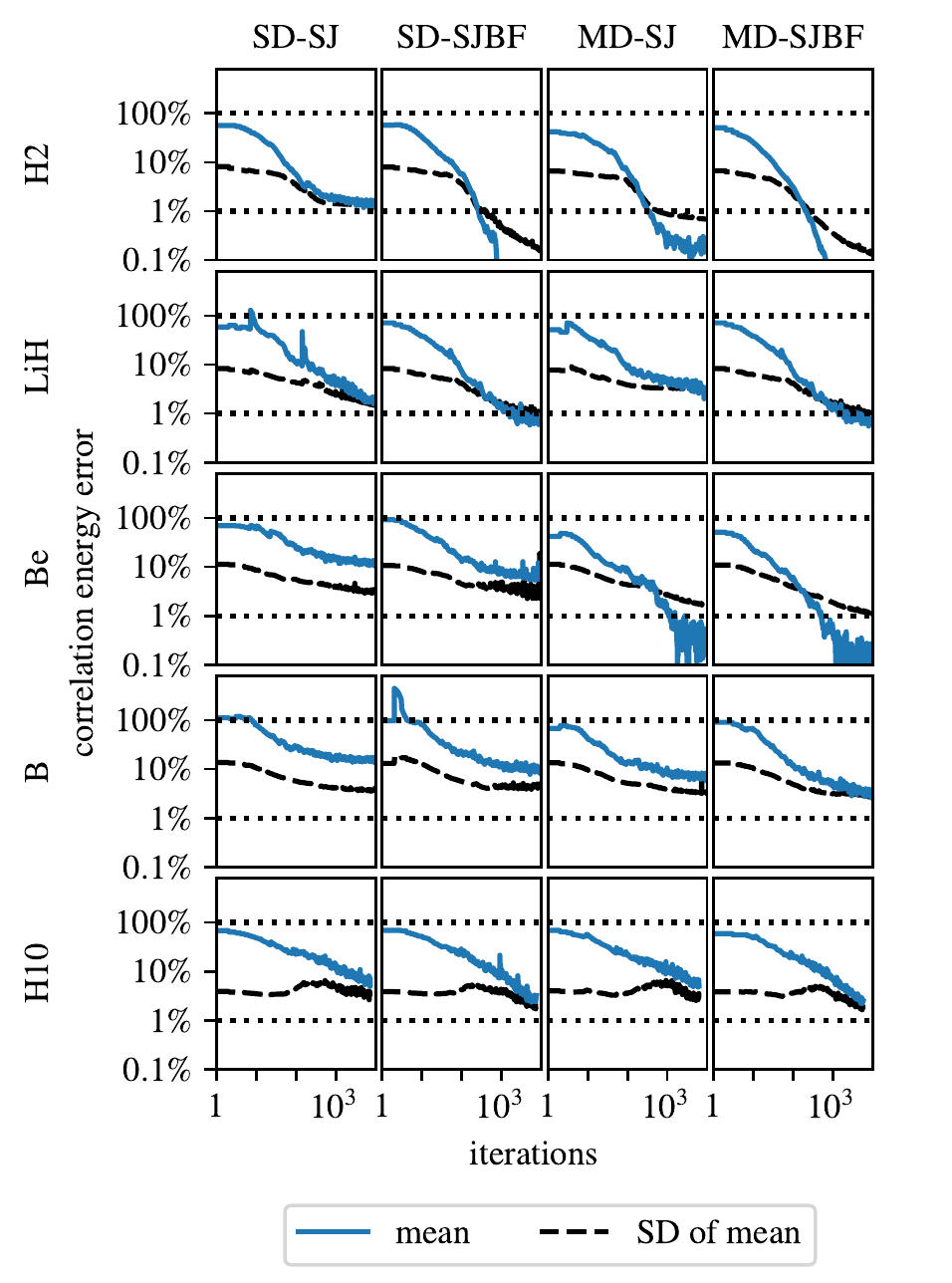}
\caption{\textbf{Learning curves of PauliNet for five molecular systems.}
The correlation energy error is calculated from the approximate ground-state energy, $E_0$, the HF energy, $E_\text{HF}$, and the exact grounds-state energy, $E_0^\text{exact}$, as $1-(E_0-E_\text{HF})/(E_0^\text{exact}-E_\text{HF})$.
The mean of the error and its estimated standard deviation ($\sigma/\sqrt{N_\text b}$) over a single batch of size $N_\text b$ are plotted for each iteration of the Adam optimization.
Optimizations of four different wave function ansatzes are shown, which are combinations of a single or multiple Slater determinants (SD/MD), a Jastrow factor (SJ), and a backflow (BF).
In the MD variants, all determinants in a normalized CASSCF expansion with a linear coefficient above 0.01 are used---H$_2$: 2 determinants, LiH\@: 6, Be: 4, B\@: 3, H$_{10}$: 36.
}\label{fig:learning-curves}
\end{figure}

We train PauliNet via the variational principle, minimizing total electronic energy (variational QMC). The training data are electron configurations that are generated on-the-fly by sampling the electron distribution, $\lvert\psi\rvert^2$ (Methods for details).
We first investigate the same systems that were used for DeepWF \citep{HanJCP19}, in particular the hydrogen molecule H$_2$, lithium hydride (LiH), beryllium (Be), boron (B), and the linear hydrogen chain H$_{10}$ (Table~\ref{tab:energies}).
For these systems, PauliNet recovers between 97\% and 99.9\% of the electron correlation energy after training for tens of minutes to few hours on a single GTX 1080 Ti graphics processing unit.
We compare these results to the standard single-determinant (SD) variational (VMC) and diffusion Monte Carlo (DMC) methods, which, unlike their multi-determinant variants, have the same asymptotic scaling as our approach.
For the studied systems, SD-VMC recovers 60\% to 92\% of the correlation energy, and SD-DMC recovers 88\% to 99.7\%, well below the accuracy of PauliNet.
The DeepWF ansatz does not even reach the baseline HF accuracy for some systems.

Figure~\ref{fig:learning-curves} highlights two crucial aspects of our results.
First, the error in the correlation energy decreases almost monotonously as the training progresses from the initial HF baseline level to the reported final values.
The learning curves in fact never plateau, demonstrating the high expressiveness of our neural network ansatz, and indicating that even higher accuracy could be reached by investing more computational resources.
Second, we compare the full ansatz to variants using only a single determinant and variants without backflow, and find that both these components are important for refining the nodal surface of the HF baseline, and thus reaching high accuracy.

Only a few determinants are sufficient to substantially reduce the correlation energy error compared to the single-determinant case: 6 or less determinants for all atoms and diatomic molecules, and 36 determinants for H$_{10}$.
Similar findings were made for FermiNet, indicating that deep learning can indeed be an efficient tool to reduce the large number of determinants sampled in other VMC approaches that directly operate on determinants of fixed orbitals \citep{BoothJCP09,ChooNC20}.



\begin{figure}[t]
\centering
\includegraphics[width=1.0\columnwidth]{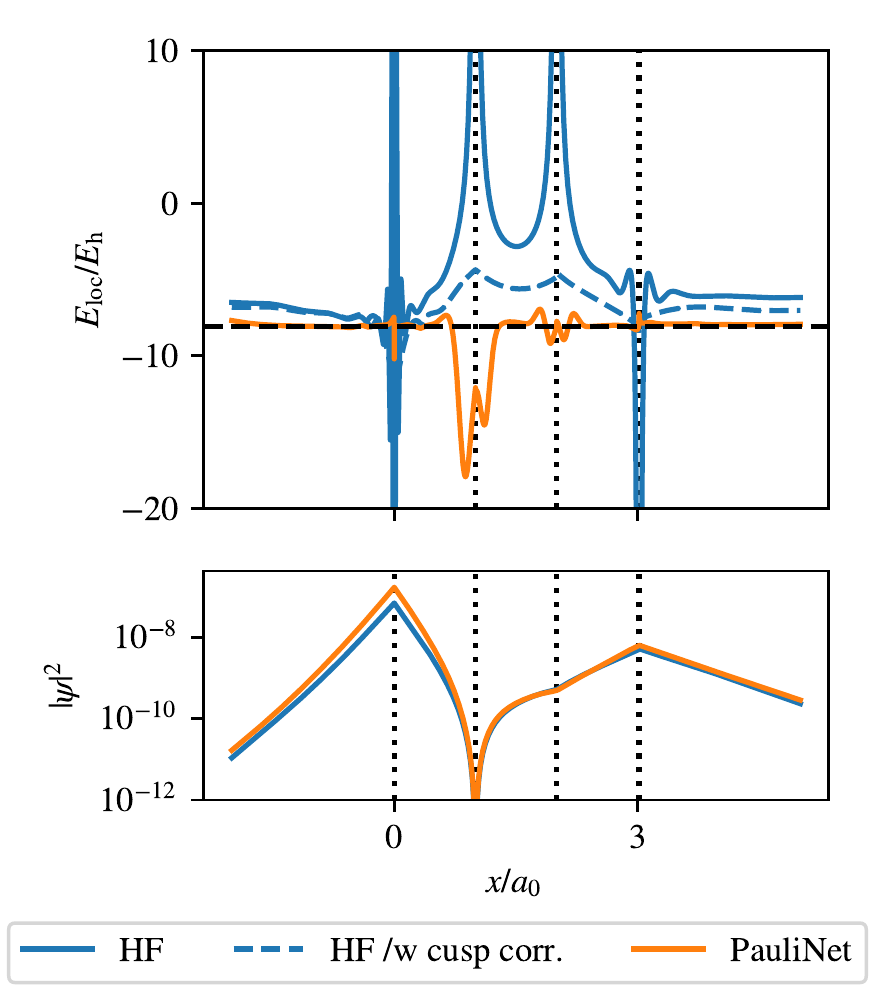}
\caption{\textbf{One-dimensional scan of the wave function and its local energy in LiH.}
Four vertical dashed lines indicate positions of fixed particles on the $x$-axis, from left the Li nucleus, a spin-up electron, a spin-down electron, and the H nucleus.
The second spin-down electron is positioned randomly off the axis, while the second spin-up electron is moved along the $x$-axis.
}\label{fig:lih-wf}
\end{figure}

The total electronic energy is a global measure of the quality of the wave function, but does not necessarily portray whether the wave function is locally correct.
Especially with the use of DNNs one could doubt whether the wave function behaves correctly in regions that have not been sampled in the training data.
Figure~\ref{fig:lih-wf} shows that PauliNet generalizes well to such regions.
Even though the chosen electron configuration has a very low probability ($\lvert\psi\rvert^2$ orders of magnitude lower than for the most likely configurations), the DNN learned to modify the HF wave function in such a way that the local energy shifts towards the exact value along the whole scan, except for a small region where the probability goes to zero due to the Pauli exclusion principle.
The plot also illustrates the correct representation of the nuclear and electronic cusps, as the local energy of the PauliNet wave function does not diverge around the coalescence points.

\subsection{Capturing strong correlation}  

\begin{figure}[t]
\centering
\includegraphics[width=1.0\columnwidth]{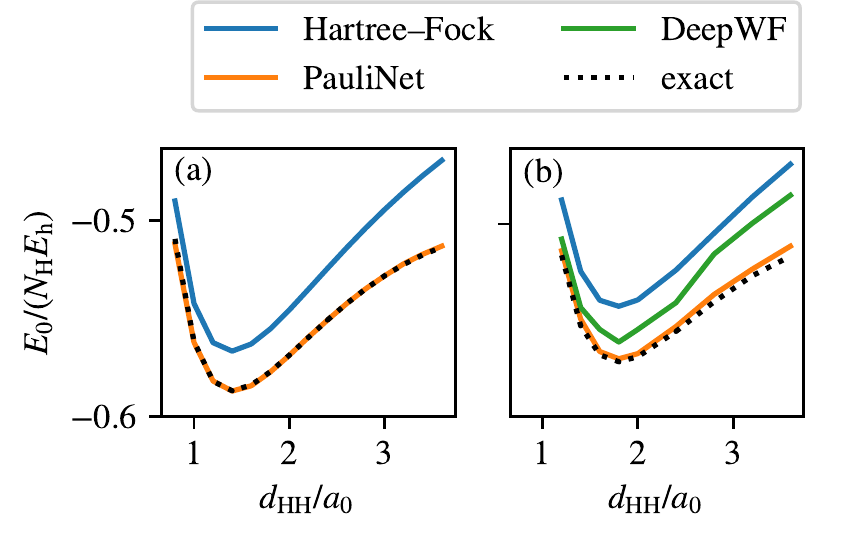}
\caption{\textbf{Electronic energy as a function of distance between hydrogen atoms in H$_2$ (a) and H$_{10}$ (b).}
The exact values are calculated with FCI@cc-pvQZ for H$_2$ and taken from \citet{MottaPRX17} for H$_{10}$.
}\label{fig:int-curves}
\end{figure}

Unlike the atoms and diatomics, the linear hydrogen chain H$_{10}$ exhibits \emph{strong correlation}, which describes a situation where the single-determinant description of the HF method is insufficient, and the correlation energy constitutes a significant part of the electronic energy \citep{MottaPRX17}.
%
Without a multi-determinant expansion, strong correlation poses a hard test for a Jastrow factor, because distinguishing the high-energy electronic configurations is a complicated function of the positions of all the electrons with respect to each other and to the nuclei, not just of the electron distances \citep{GoetzJCTC17}.
Already H$_2$ exhibits strong correlation when the two atoms are dissociated, and PauliNet is able to  recover essentially 100\% of the correlation energy along the whole dissociation curve (Figure~\ref{fig:int-curves}a), even in the stripped-down single-determinant form.
For H$_{10}$, we recover 98\% (90\%) of the exact correlation energy in the equilibrium (stretched) geometry (Figure~\ref{fig:int-curves}b) with the full multi-determinant form.
The decrease of the captured fraction of the correlation energy with larger atom separations illustrates the increased difficulty of the problem due to the strong correlation.
Even with a single determinant, our deep Jastrow factor is able to recover more than 90\% of the electron correlation for the equilibrium geometry, demonstrating its flexibility.

\section{Discussion}  

We have designed PauliNet, a DNN representation of electronic wave functions in real space and shown that it can outperform state-of-the-art variational quantum chemistry methods that do not use large determinant expansions. 
In contrast, our approach requires only a few determinants, and as a result we anticipate that its computational cost scales asymptotically as $N^4$ ($N^3$ for a determinant evaluation, and additional $N$ for the evaluation of the kinetic energy)
PauliNet is thus a candidate for a quantum chemistry method that can scale to much larger systems with high accuracy.

Compared to standard functional forms used in QMC, the use of DNNs has several advantages.
First, the much higher flexibility of DNNs allows a variational approach to reach or exceed the accuracy of  diffusion QMC, which facilitates the calculation of accurate derived electronic properties beyond the electronic energy.
Second, besides encoding more complex many-body correlations between electrons, DNNs have an essentially unlimited flexibility in the spatial degrees of freedom, circumventing the curse of incomplete basis sets of quantum chemistry, which can be removed only with DMC when using standard techniques.
Third, in standard QMC, the wild oscillations of the local energy close to heavy nuclei mandates the use of pseudopotentials for heavier elements such as transition metals. The flexibility of DNNs could sidestep this necessity by smoothing out these oscillations, which cannot be done with standard wave-function ansatzes.

In classical quantum chemistry methods, strong correlation is usually treated by using large multi-determinant expansions, which are computationally demanding and introduce the problem of selecting the proper subset of determinants.
Treating strong correlation on the level of Jastrow factors traditionally requires construction of specialized many-body forms \citep{GoetzJCTC17}.
In contrast, DNNs are capable of learning strong correlation between electrons without any specialized adaptation.
We demonstrate convergence to very high accuracy with single or a few determinants, changing the problem from searching or sampling over exponentially many determinants to letting a deep neural network search over exponentially many functions.
Although it is unclear whether this is advantageous in a strict mathematical sense, this is precisely the task which deep neural networks have been demonstrated to be strong at in a variety of real-world applications.
Complementary approaches that use variational QMC in a second-quantized form of the electronic problem have also been proposed \citep{ChooNC20}.
This class of methods has the advantage of eliminating much of the complexity of electronic wave functions, such as the antisymmetry or cusp conditions, from the machine-learning part of the problem, but needs to cope with the ubiquitous limitations of single-particle basis sets.

In parallel, \citet{Pfau19} developed FermiNet, a different deep-learning architecture to solve the electronic Schrödinger equation.
Both methods achieve nearly 100\% of the correlation energy for the systems tested, illustrating the versatility of deep QMC approaches. The residual error of FermiNet is even smaller than that of PauliNet, but at the expense of computational cost that is approximately two orders of magnitude greater.
Lacking learning curves similar to our Figure~\ref{fig:learning-curves} for FermiNet prevents any detailed comparison of the computational efficiency at this point.
However, the results reported by \citet{Pfau19} were obtained with about 30 times more electronic samples for training, using eight state-of-the-art graphics processing units in parallel instead of one in our case.
Establishing a shared protocol for evaluating the accuracy per computational effort for DNN wave function ansatzes will be a an important task in order to judge efficiency and usefulness in the future.

Already a brief comparison of the two approaches hints at potential improvements of both architectures.
The combination of architectural design and optimization methods used in FermiNet with the built-in physical constraints of PauliNet appears to be a promising venue for computationally affordable, scalable, yet highly accurate black-box methods for quantum chemistry.
We hope that the introduction of neural networks into the field of electronic QMC opens the possibility to utilize the striking advances in deep learning from the last decade in a new field.

\section{Methods}  

\paragraph{Ansatz optimization}  

We optimize the PauliNet ansatz individually for each atomic system in an unsupervised fashion using the variational principle for the total electronic energy,
\begin{equation}
\begin{gathered}
  E_0=\min_\psi E[\psi]\leq\min_{\boldsymbol\theta}E[\psi_{\boldsymbol\theta}],\\
  E[\psi]=\int\mathrm d\mathbf r\,\psi(\mathbf r)\hat H\psi(\mathbf r)
\end{gathered}
\end{equation}
%
Following the standard QMC technique, the energy integral is evaluated as an expected value of the \emph{local energy}, $E_\text{loc}[\psi](\mathbf r)=\hat H\psi(\mathbf r)/\psi(\mathbf r)$, over the probability distribution $\lvert\psi^2(\mathbf r)\rvert$,
\begin{equation}
  E[\psi]=\mathbb E_{\mathbf r\sim|\psi|^2}\big[E_\text{loc}[\psi](\mathbf r)\big]
  \label{eq:monte-carlo}
\end{equation}
We generate training data for PauliNet on-the-fly by sampling electron positions
with a standard Langevin Monte Carlo approach~\citep{UmrigarJCP93} every 100 training iterations using the current model wave function $\psi_{\boldsymbol\theta}$.
Each sampled electron configuration is used only once in an optimization run.
We use a simplified version of the method by \citet{UmrigarJCP93}, in which the radial step proposal is replaced with clipping the step length such that the step size is always shorter than the distance to the nearest nucleus, so the nucleus can never be ``overshot''.
The initial electron positions for the Markov chain are sampled from Gaussian distributions around the nuclei such that the effective atomic Mulliken charges obtained from the HF method are respected.

To optimize the parameters $\boldsymbol\theta$ in the Jastrow and backflow neural networks, we use the weighted Adam optimizer \citep{Kingma17,LoshchilovICLR19} together with the total energy used directly as the loss function.
To calculate the stochastic gradient of the loss function over a batch of samples, we use a gradient formula that takes advantage of the fact that the Hamiltonian operator is Hermitian~\citep{CeperleyPRB77},
\begin{equation}
\begin{gathered}
\mathcal L(\boldsymbol\theta)
  =\mathbb E_{\mathbf r\sim\lvert\psi'^2\rvert}
  \big[E_\text{loc}[\psi_{\boldsymbol\theta}](\mathbf r)\big] \\
\boldsymbol\nabla_{\boldsymbol\theta}\mathcal L(\boldsymbol\theta)
  =2\mathbb E_{\mathbf r\sim\lvert\psi'^2\rvert}\big[
    \big(E_\text{loc}[\psi_{\boldsymbol\theta}](\mathbf r)
    -\mathcal L(\boldsymbol\theta)\big)
    \boldsymbol\nabla_{\boldsymbol\theta}\ln\lvert\psi_{\boldsymbol\theta}\rvert
  \big]
\end{gathered}\label{eq:trainig}
\end{equation}
This expression for the gradient requires calculating only second derivatives of the wave function (for the Laplace operator), whereas direct differentiation would require third derivatives (derivative of the Laplace operator), which is computationally costly and numerically unstable.
We smoothly clip the local energy of each sample by a logarithmically growing clipping function outside the window defined as 5 times the mean deviance from the median local energy in a given batch.

\paragraph{Cusp conditions}  


Eq.~\ref{eq:ansatz} ensures the nuclear cusp conditions via the molecular orbitals $\varphi_\mu(\mathbf r_i)$.
We achieve this by modifying the molecular orbitals using the technique from \citet{MaJCP05} with one simplification---we optimize the orbital values at atomic nuclei, $\mathbf r_i=\mathbf R_I$, via the energy variational principle, rather then fitting them against references values.
The electronic cusp conditions are enforced by $\gamma(\mathbf r)$,
\begin{equation}
\gamma(\mathbf r):=\sum_{i<j}-\frac {c_{ij}}{1+\lvert\mathbf r_i-\mathbf r_j\rvert},
\end{equation}
where $c_{ij}$ is either $\tfrac12$ or $\tfrac14$ depending on the spins of the two electrons.
To preserve the cusp conditions built into $\varphi_\mu$ and $\gamma$, the Jastrow factor and backflow DNNs must be cuspless,
\begin{equation}
\nabla_{\mathbf r_i}J(\mathbf r)
  \big\rvert_{\mathbf r_i=\{\mathbf r_k,\mathbf R_a\}}=0,\quad
\nabla_{\mathbf r_i}f_{\mu i}(\mathbf r)
  \big\rvert_{\mathbf r_i=\{\mathbf r_k,\mathbf R_a\}}=0
  \label{eq:cuspless}
\end{equation}
These conditions are ensured by construction by the DNNs, as detailed below.

\paragraph{PauliNet extension of SchNet}  

SchNet is an instance of the class of graph convolutional neural networks, and was designed to model the electronic energy as a function of just the nuclear charges and coordinates \citep{SchuttJCP18}.
In PauliNet, we use SchNet to represent electrons in molecular environments by implementing the iteration rule in Eq.~\ref{eq:schnet-iter},
\begin{equation}
\begin{aligned}
\mathbf z_i^{(n,\pm)}&:=\sum\nolimits_{j\neq i}^\pm
  \mathbf w^{(n,\pm)}_{\boldsymbol\theta}
  \big(\mathbf e(\lvert\mathbf r_i-\mathbf r_j\rvert)\big)
  \odot\mathbf h_{\boldsymbol\theta}^{(n)}\big(\mathbf x_j^{(n)}\big) \\ 
\mathbf z_i^{(n,\mathrm n)}&:=\sum\nolimits_I
  \mathbf w_{\boldsymbol\theta}^{(n,\mathrm n)}
  \big(\mathbf e(\lvert\mathbf r_i-\mathbf R_I\rvert)\big)
  \odot\mathbf Y_{\boldsymbol\theta,I} \\
\mathbf x_i^{(n+1)}&:=\mathbf x_i^{(n)}
  +\sum\nolimits_\pm\mathbf g^{(n,\pm)}_{\boldsymbol\theta}
  \big(\mathbf z_i^{(n,\pm)}\big)
  +\mathbf g^{(n,\mathrm n)}_{\boldsymbol\theta}
  \big(\mathbf z_i^{(n,\mathrm n)}\big)
\end{aligned}
\end{equation}
where ``$\odot$'' denotes element-wise multiplication, $\mathbf w_{\boldsymbol\theta}^{(n)}$, $\mathbf h^{(n)}_{\boldsymbol\theta}$, and $\mathbf g_{\boldsymbol\theta}^{(n)}$ are trainable functions represented by ordinary fully-connected DNNs, and $\mathbf e$ is a function that featurizes the interatomic distances.
The modifications of the original SchNet are as follows.
\begin{enumerate}[label=(\roman*),itemsep=0pt]  
\item Since the wave function is a function of electron coordinates, the iterated feature vectors $\mathbf x_i^{(n)}$ represent electrons, not atoms.
\item The messages $\mathbf z_i(n)$ received by the electron feature vectors at each iteration are split into three channels, corresponding to same-spin electrons (+), opposite-spin electrons ($-$), and the nuclei (n).
This builds more flexibility into the architecture, and is motivated by the fact electrons and nuclei are particles of entirely different type.
\item Each channel has a separate receiving function $\mathbf g_{\boldsymbol\theta}$, again increasing flexibility without substantially increasing the number of parameters.
\item Each nucleus is represented by a trainable embedding $\mathbf Y_{\boldsymbol\theta,I}$, which is shared across all iterations and not iteratively updated.
In VMC, the wave function is always optimized for a given fixed geometry of the nuclei, so the nuclear embeddings can be assumed to already represent each nucleus with its (fixed) atomic environment, hence the absence of need for their iterative refinement.
\item The distance features~$\mathbf e$ are constructed to be cuspless, as detailed below.
\end{enumerate}
We use a distance featurization inspired by the PhysNet architecture \citep{UnkeJCTC19}, with a modified envelope that forces all the Gaussian features and their derivatives to zero at zero distance,
\begin{align}
  e_k(r)   &:=r^2\mathrm e^{-r-{(r-\mu_k)}^2/\sigma_k^2}, \\
  \mu_k    &:=r_\text cq_k^2 \\
  \sigma_k &:=\frac{1}{7}(1+r_\text cq_k)
\end{align}
where $q_k$ equidistantly spans the interval $(0,1)$ and $r_\text c$ is a cutoff parameter.



\paragraph{Computational details}  

All reported methods were implemented with Pytorch in the open-source DeepQMC package, which is available on Zenodo \citep{DeepQMC} and developed on Github at \url{https://github.com/deepqmc/deepqmc}.
The linear coefficients of the HF orbitals $\varphi_\mu$ as well as of the determinants in a multideterminant expansion were calculated with PySCF \citep{SunWCMS18} using the ``6-311G'' Gaussian basis set.  
The plain fully-connected DNNs that represent the trainable functions in our architecture were chosen such the total number of trainable parameters is around \num{7e4} (see Table~\ref{tab:hyper}).

\begin{table}[h]
\centering
\caption{\textbf{Hyperparameters used in numerical calculations.}
}\label{tab:hyper}
\begin{tabular}{lc}
\toprule
Hyperparameter & Value \\
\midrule
One-electron basis & 6-311g \\  
Dimension of $\mathbf e$ & 32 \\
Dimension of $\mathbf x_i$ & 128 \\
Dimension of $\mathbf z_i$ & 64 \\
Number of interaction layers $L$ & 3 \\
Number of layers in $\eta_{\boldsymbol\theta}$ & 3 \\
Number of layers in $\boldsymbol\kappa_{\boldsymbol\theta}$ & 3 \\
Number of layers in $\mathbf w_{\boldsymbol\theta}$ & 2 \\
Number of layers in $\mathbf h_{\boldsymbol\theta}$ & 1 \\
Number of layers in $\mathbf g_{\boldsymbol\theta}$ & 1 \\
Determinant cutoff & \num{1e-2} \\
Learning rate & 0.01 \\
Batch size & \num{10000} \\
Number of walkers & \num{2000} \\
Number of training steps & \num{7000} \\
Optimizer & AdamW \\
Learning rate decay period $t_0$ & 200 \\
Clipping window $q$ & 5 \\
Resampling period & 100 \\
Number of discarded sampling steps & 50 \\
Number of decorrelation sampling steps & 1 \\
Target acceptance & 57\% \\
\bottomrule
\end{tabular}
\end{table}

\begingroup
\setlength\bibsep{0pt}
\newcommand{\mathsl}{\mathit}
\footnotesize
\bibliography{refs-zotero,refs}
\endgroup

\subsection*{Acknowledgments}

\begingroup
\footnotesize
We are grateful to the following scientists for inspiring discussions:
Cecilia Clementi (Rice, FU Berlin), Jens Eisert (FU Berlin), Gustavo
Scuseria (Rice), Jörg Neugebauer (MPIE), and Hao Wu (Tongji). Funding is
acknowledged from the European Commission (ERC CoG 772230
``Scale-Cell''), Deutsche Forschungsgemeinschaft (CRC1114/A04, GRK2433
DAEDALUS/P04), the MATH\(^+\) Berlin Mathematics research center (AA1x6,
EF1x2). J.~H.~thanks K.-R.~Müller for support and acknowledges funding
from TU Berlin (Project No.~10032745).

\endgroup

%

\end{document}